\documentclass[runningheads]{llncs}
\usepackage{booktabs}    
\usepackage{array}  
\usepackage{tabularx}  
\usepackage[T1]{fontenc}
\usepackage{graphicx}
\begin{document}
\title{Re-Sonance: A Dysarthric Asynchronous Real-Time Speech Conversion System Based on a Three-Stage Cascaded ASR-LLM-TTS Architecture}
\titlerunning{A Dysarthric Asynchronous Real-Time Speech Conversion System}
%
\author{Yuxuan Wu \and Yifan Xu \and Junkun Wang \and Jiayong Jiang \and Xin Zhao \and
Zhaojie Luo\thanks{Corresponding author, email: luozhaojie@seu.edu.cn}}
\authorrunning{Wu et al.}
\institute{State Key Laboratory of Digital Medical Engineering, Southeast University, Nanjing, China
\and
School of Biological Science \& Medical Engineering, Southeast University, Nanjing, China}

\maketitle              
\begin{abstract}
Individuals with dysarthria face significant challenges in professional speaking scenarios such as conferences, presentations, and meetings, where real-time communication is crucial. While existing Augmentative and Alternative Communication (AAC) systems provide basic support, they often fail to meet the demands of professional speaking environments due to high latency and unnatural speech patterns. This paper presents Re-Sonance, a novel LLM-enhanced speech-driven AAC system designed for real-time professional speaking scenarios. By integrating Whisper ASR, Qwen LLM, and CosyVoice TTS, Re-Sonance achieves improved speech intelligibility and naturalness while maintaining real-time performance. Both subjective and objective evaluations using a Mandarin dysarthric speech dataset demonstrate that our speech reconstruction approach significantly improved intelligibility while preserving semantic coherence for speakers with mild to moderate dysarthria. Although performance remains limited for severe dysarthria cases, our findings validate the potential of LLM-based methods for enhancing speech-driven AAC systems, paving the way for more effective and accessible communication technologies.

\end{abstract}

\section{Introduction}
Dysarthria, a group of congenital or trauma-induced neuromotor disorders, impairs articulation and communication\cite{rudzicz2010articulatory}, significantly affecting patients' linguistic abilities, social interactions, and quality of life. Although advances in Augmentative and Alternative Communication (AAC) technologies have improved expression for individuals with dysarthria, recent research has largely emphasized speech replacement, aiming to reduce user input and streamline workflows\cite{valencia2023less}. However, these methods disrupt natural communication patterns and face limitations in comprehensibility and real-time performance.

The unnatural interaction imposed by speech replacement can undermine users' self-identity, social recognition, and motivation to use assistive systems\cite{ayoka2024enhancing,ibrahim2018design}. Additionally, their operational complexity and latency hinder use in real-time settings like video conferences and public speaking. To address these issues, we propose an AAC system tailored for real-time meetings and presentations, enabling natural participation through intuitive, minimally intrusive interaction.

Building on recent models, our system ensures broad accessibility across devices and languages. The preprocessing module uses lightweight speech recognition for initial interpretation of dysarthric speech, followed by large language models (LLMs) to refine output and mitigate recognition errors through prompt engineering. The output module employs lightweight text-to-speech (TTS) to produce natural, intelligible speech.

This work bridges key gaps in AAC systems and advances speech-driven assistive communication, contributing to global efforts toward communication equity for individuals with disabilities.

\section{Related Works}
\subsection{Human-Computer Interaction for Individuals with Disabilities}
Research on Augmentative and Alternative Communication (AAC) has long explored tools such as eye-tracking\cite{cai2023speakfaster} and touch-based keyboards with features like word prediction and phrase storage\cite{mitchell2022ability,valencia2024compa,mohan2024powerful}. These are effective for severe impairments but less suitable for dysarthria, as replacing speech reduces communication speed\cite{kane2017times} and may undermine self-identity or rehabilitation opportunities\cite{dai2022designing}.
\subsection{Speech-Driven AAC Systems}
Speech provides a natural, high-bandwidth channel\cite{munteanu2018speech}. Existing work falls into two main approaches: signal processing and ASR-TTS frameworks\cite{zheng2023comparing}.
\subsubsection{Signal Processing Methods}
Deep learning methods enhance or convert speech, e.g., accent\cite{jia2024convert}, emotion\cite{luo2022decoupling,qi2024pavits}, or dysarthric-to-healthy patterns\cite{chu2023dgan,wang2024unit}. Tools like Wesper convert whispered to normal speech\cite{rekimoto2023wesper}. However, these approaches often demand high resources and may not improve intelligibility in practice.
\subsubsection{ASR-TTS Framework}
ASR-TTS systems are widely applied in AAC\cite{wu2024finding,kim2021can}, but inconsistent ASR errors remain a barrier\cite{hong2018identifying,tran2024assessment}. Personalized systems like Google’s Project Relate address this\cite{ayoka2024enhancing}, though challenges remain in language coverage and domain-specific terms\cite{deshpande2014crowdsourcing}.
\subsection{LLM-Driven Assistive Systems}
Large Language Models (LLMs) enable new assistive applications. Rambler summarizes ASR outputs\cite{lin2024rambler}, AscleAI refines clinical transcripts\cite{han2024ascleai}, and Valencia et al. show improved AAC interaction via speech macros\cite{valencia2020conversational}. These works highlight LLMs’ potential to correct ASR errors and enhance clarity in real-time communication.

\section{Method}
\subsection{Re-Sonance Prototype}
\begin{figure}[h]
  \centering
  \includegraphics[width=0.9\linewidth]{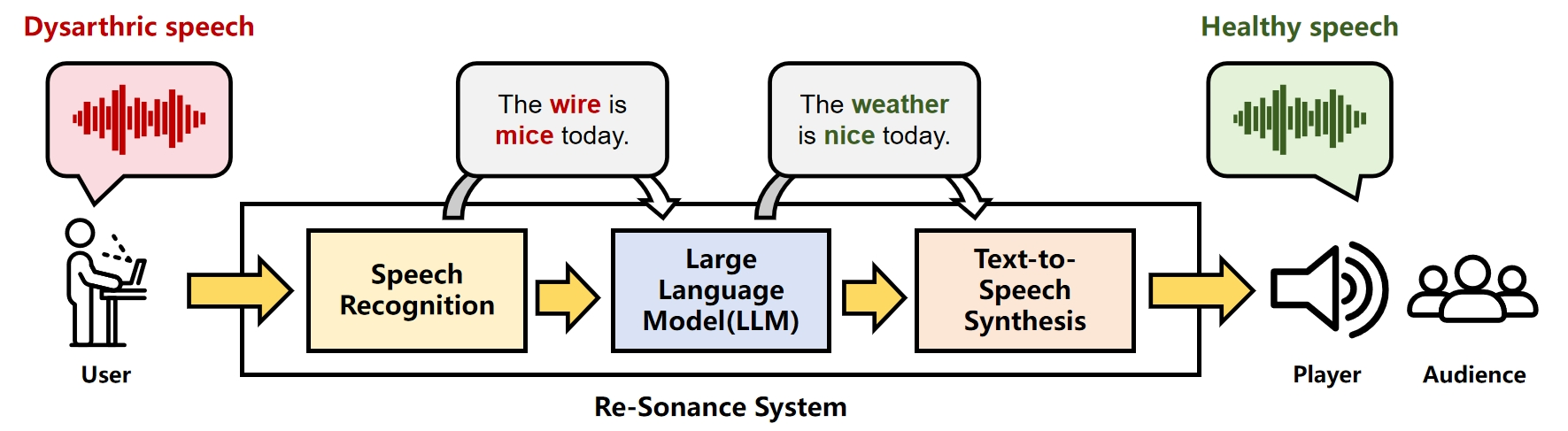}
  \caption{Technical framework and application scenarios of Re-Sonance}\label{fig_sys}
\end{figure}

The Re-Sonance system is designed to convert pathological speech from individuals with dysarthria into healthy speech. The technical framework of Re-Sonance consists of three primary components (Fig.~\ref{fig_sys}), which function collaboratively to transform dysarthric speech into more intelligible output.

For speech recognition, we employ Whisper, one of the most advanced open-source ASR models developed by OpenAI\cite{radford2023robust}. Released in 2022, Whisper supports 99 languages and can be deployed on low-performance devices, demonstrating robustness and accuracy in recognizing healthy speech. While some fine-tuned versions of Whisper exist for dysarthric speech, this study utilizes the unmodified Whisper-Turbo model.
For the large language model, we use Qwen-Plus, an enhanced large-scale language model developed by Alibaba Cloud\cite{bai2023qwen}. Qwen-Plus excels in language comprehension and generation, supports multiple languages, and optimally balances performance, speed, and cost, making it suitable for diverse applications.
For TTS synthesis, we employ CosyVoice, an advanced speech synthesis tool that generates natural-sounding speech with support for multiple languages and personalized adjustments\cite{du2024cosyvoice}. Additionally, CosyVoice offers voice cloning capabilities, enabling highly realistic speech synthesis from a few seconds of audio samples.

\subsection{Asynchronous Real-Time Processing}
To achieve low-latency interaction, Re-Sonance adopts an asynchronous real-time design. 
Here, "asynchronous" refers to the non-blocking pipeline architecture, where different modules 
(ASR, LLM, and TTS) operate in a streaming and overlapping manner rather than strictly sequential execution. 
Specifically, the Whisper ASR continuously transcribes dysarthric speech into partial text segments. These segments are immediately passed to the Qwen LLM for correction and refinement, without waiting for the entire utterance to finish. In turn, the corrected segments are incrementally fed into CosyVoice TTS, which can synthesize speech outputs on-the-fly. 
This overlapping process effectively reduces end-to-end latency compared with synchronous pipelines, ensuring that the user experiences smooth and near-instantaneous communication. 

\subsection{Interface}
\begin{figure}
  \centering
  \includegraphics[width=0.85\textwidth]{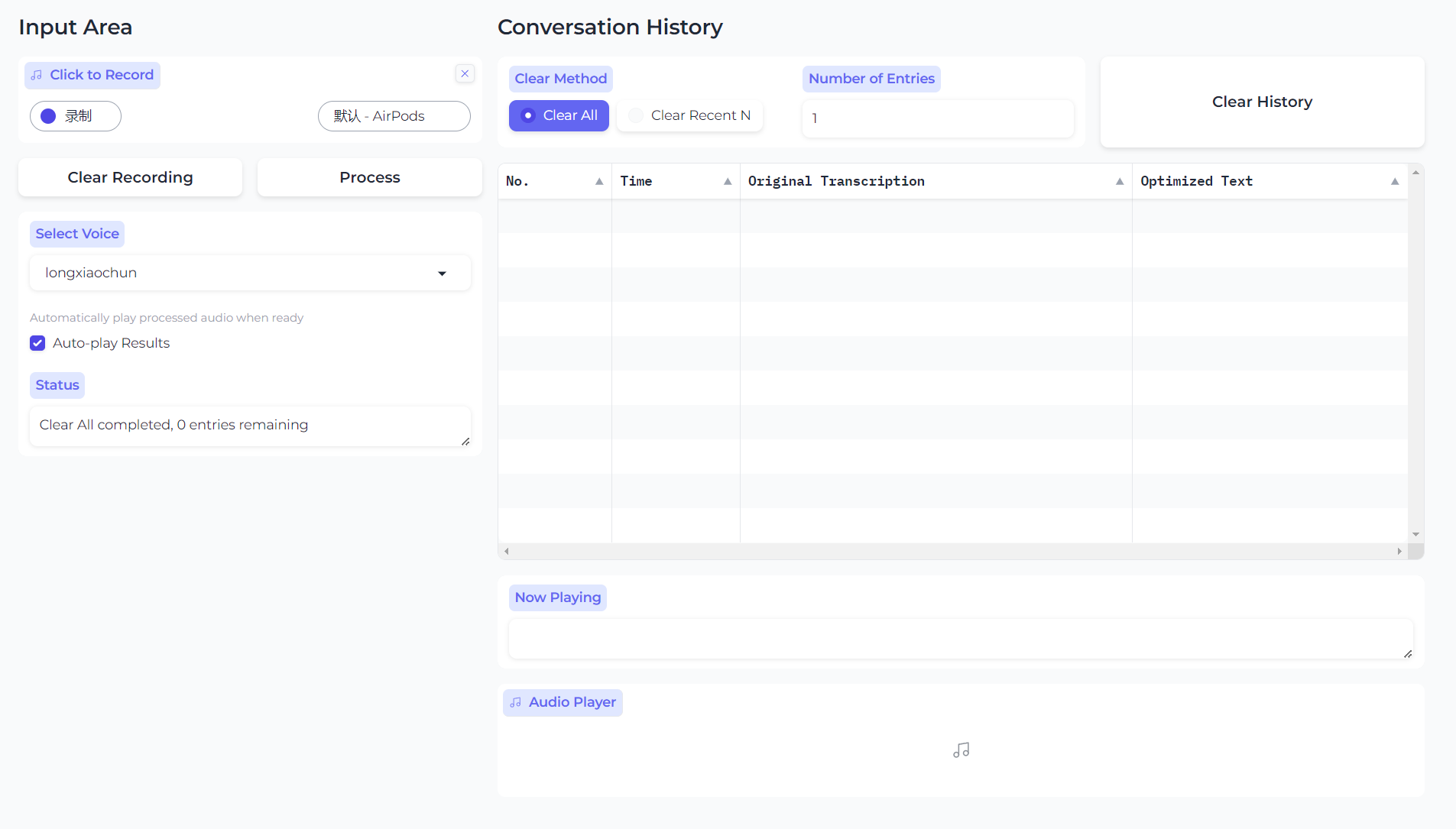}
  \caption{Interface of the Re-Sonance}\label{fig_interface}
\end{figure}

The Re-Sonance interface consists of input and output control components, optimized for both mobile and desktop platforms\footnote{We have created a demo: \url{https://demo-resonance.hai-lab.cn/}.}. Fig.~\ref{fig_interface} presents the user interface of the Re-Sonance system.

\subsection{Experimental Procedure}
To investigate the potential of LLM-based speech-driven AAC systems in dysarthric speech recognition, we conducted simulation experiments using the Chinese Dysarthric Speech Dataset (CDSD)\cite{wan2024cdsd}. The evaluation comprised both subjective and objective assessments.
\subsubsection{Subjective Evaluations}
Speech samples from 10 individuals with varying degrees of dysarthria were selected from the CDSD dataset to reflect real-world conditions. Twenty native Chinese speakers participated in the evaluation, assessing three dimensions: intelligibility, naturalness, and semantic relevance—the last of which measures the alignment between generated speech and the speaker’s original intent. The evaluation covered three types of speech: original dysarthric speech, baseline ASR-TTS output, and LLM-enhanced ASR-TTS speech (Re-Sonance system).
As shown in Table 1, participants reviewed a slide presentation with embedded audio samples and completed an online questionnaire. All evaluators underwent prior online training with annotated scoring examples to ensure rating consistency.
\begin{table}[h]
  \caption{Rating Scale for Subjective Evaluation}
  \label{tab:commands}
  \centering
  \begin{tabularx}{\textwidth}{c X X}
    \toprule
    Score & Intelligibility Description & Naturalness Description \\
    \midrule
    5 & Fully intelligible, all information is clear & Highly natural, indistinguishable from human speech \\
    4 & Mostly intelligible, minor blurring & Generally natural, minor discontinuities \\
    3 & Partially intelligible, significant effort needed & Moderately natural, noticeable issues \\
    2 & Difficult to understand, most information unclear & Less natural, significant quality issues \\
    1 & Completely unintelligible & Highly unnatural, severely distorted \\
    \bottomrule
  \end{tabularx}
\end{table}

\subsubsection{Objective evaluations}
The objective assessment focused on transcription accuracy and latency measurements of the Re-Sonance system, evaluating LLM optimization effectiveness using CDSD samples.
For accuracy evaluation, speech samples from 30 speakers were categorized into mild, moderate, and severe dysarthria groups. Accuracy metrics included Word Error Rate (WER), Match Error Rate (MER), and Word Information Lost (WIL), analyzed across severity levels.
Latency evaluation involved processing 200 approximately 10-second clips through Re-Sonance, measuring ASR latency, LLM optimization latency, and speech generation latency. Additionally, we computed the ratio of total latency to speech duration, a crucial metric for clip-based interaction in the Re-Sonance system.
\section{Result}
\subsection{Subjective Evaluation}
\begin{figure}[h]
\centering
\includegraphics[width=\linewidth]{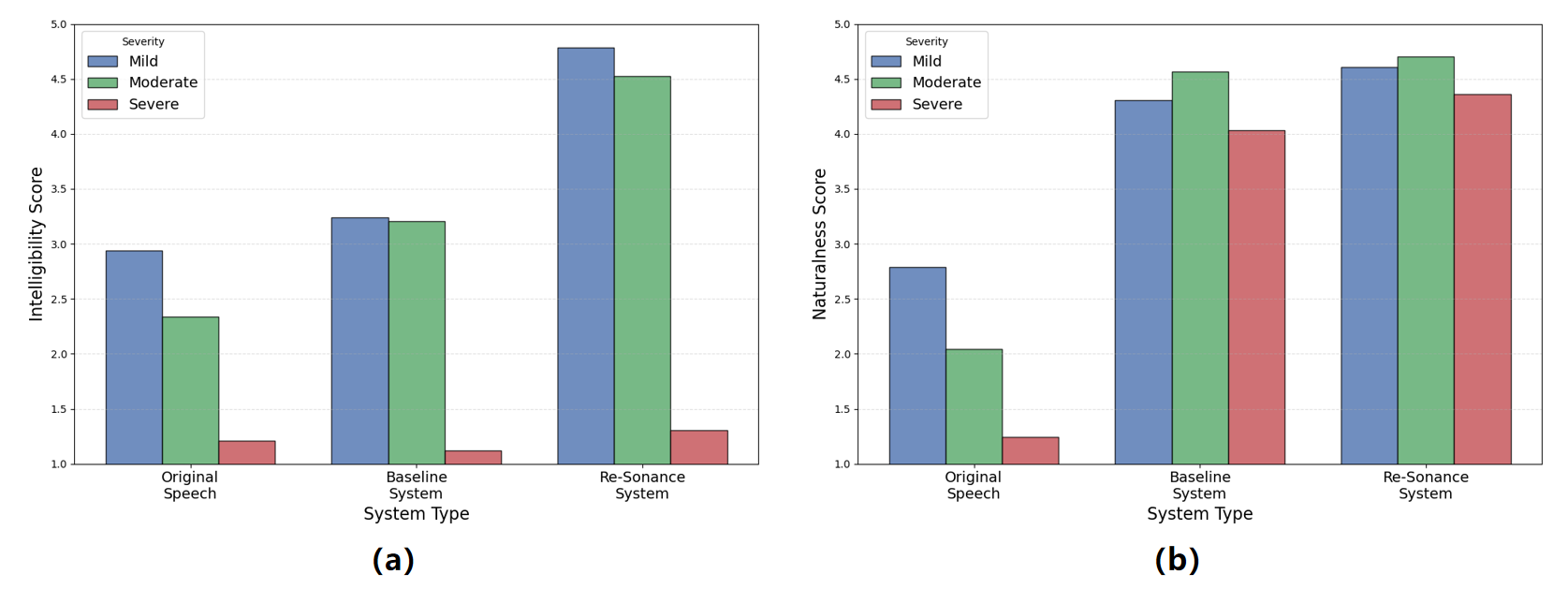}
\caption{Distribution of (a) Intelligibility and (b) Naturalness scores across system types} \label{fig_ressb}
\end{figure}

We recruited 20 native Mandarin speakers to evaluate the Re-Sonance system. Each participant rated 10 speech samples from the CDSD dataset on intelligibility, naturalness, and semantic association (Fig.~\ref{fig_ressb}).

For mild articulation disorders (n=3), Re-Sonance showed significant improvements in intelligibility (M=4.79, SD=0.48) and naturalness (M=4.61, SD=0.95) versus baseline (M=3.24, SD=0.92; M=4.30, SD=0.63). The semantic association ratio increased from 78.8\% to 98.3\%.
With moderate disorders (n=4), the system achieved higher intelligibility (M=4.55, SD=0.97) and naturalness (M=4.70, SD=0.59) compared to baseline (M=3.14, SD=1.02; M=4.57, SD=0.62), with semantic association improving from 68.2\% to 88.6\%.
However, for severe disorders (n=3), improvements were limited. Re-Sonance showed minimal gains in intelligibility (M=1.24, SD=0.60) and naturalness (M=4.36, SD=0.48) versus baseline (M=1.09, SD=0.51; M=4.03, SD=0.67). Semantic association remained low, increasing only from 3.0\% to 6.1\%, due to inherent speech recognition challenges with severely impaired speech.
\subsection{Objective evaluation}
\subsubsection{Transcription accuracy}
\begin{figure}[h]
\includegraphics[width=\textwidth]{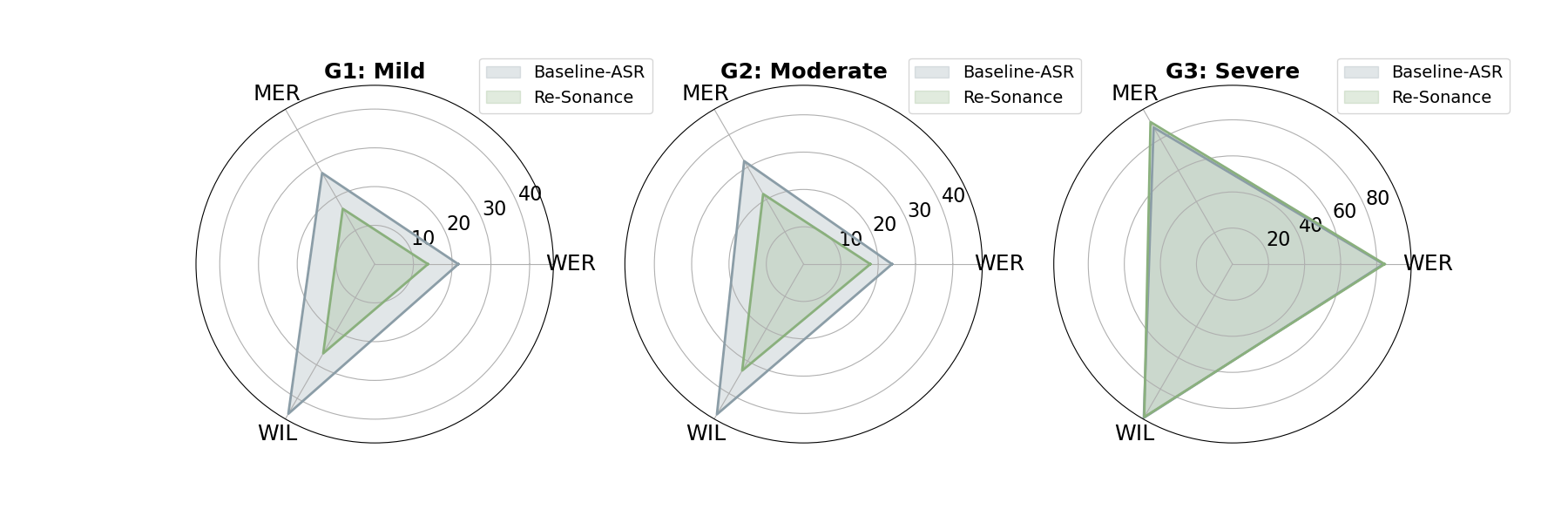}
\caption{Comparison of Baseline-ASR and Re-Sonance for Different Severity Levels} \label{fig_resob}
\end{figure}

The Re-Sonance system yielded notable improvements in speech recognition metrics for mild (G1) and moderate (G2) dysarthria groups (Fig.~\ref{fig_resob}). For G1, relative improvements were observed in WER (-7.84\%), MER (-10.66\%), and WIL (-18.00\%). Similar enhancements were achieved in G2, with improvements in WER (-5.82\%), MER (-10.18\%), and WIL (-13.54\%).
However, for patients with severe dysarthria (G3), the system exhibited suboptimal performance, as evidenced by increases in WER (+0.63\%) and MER (+3.46\%). Only WIL shows slight optimization (-0.51\%). Changes in speech recognition accuracy indicators after speech reconstruction for various levels of dysarthria are shown in Table 2.

\begin{table}[h]
  \caption{Transcription Accuracy Results of the Re-Sonance System}
  \label{tab:objective_evaluation}
  \centering
  \begin{tabular}{l c ccc ccc ccc}
    \toprule
    Severity Level & Count & \multicolumn{3}{c}{WER} & \multicolumn{3}{c}{MER} & \multicolumn{3}{c}{WIL} \\
    \cmidrule(lr){3-5} \cmidrule(lr){6-8} \cmidrule(lr){9-11}
     & & ASR & Re-Son. & $\Delta$ & ASR & Re-Son. & $\Delta$ & ASR & Re-Son. & $\Delta$ \\
    \midrule
    G1: Mild & 10 & 21.58 & 13.74 & -7.84 & 27.14 & 16.48 & -10.66 & 44.59 & 26.59 & -18.00 \\
    G2: Moderate & 10 & 23.70 & 17.88 & -5.82 & 31.87 & 21.69 & -10.18 & 46.47 & 32.93 & -13.54 \\
    G3: Severe & 10 & 83.77 & 84.40 & +0.63 & 87.50 & 90.96 & +3.46 & 98.40 & 97.89 & -0.51 \\
    \bottomrule
  \end{tabular}
\end{table}
\subsubsection{Latency}
\begin{figure}[h]
\includegraphics[width=0.9\linewidth]{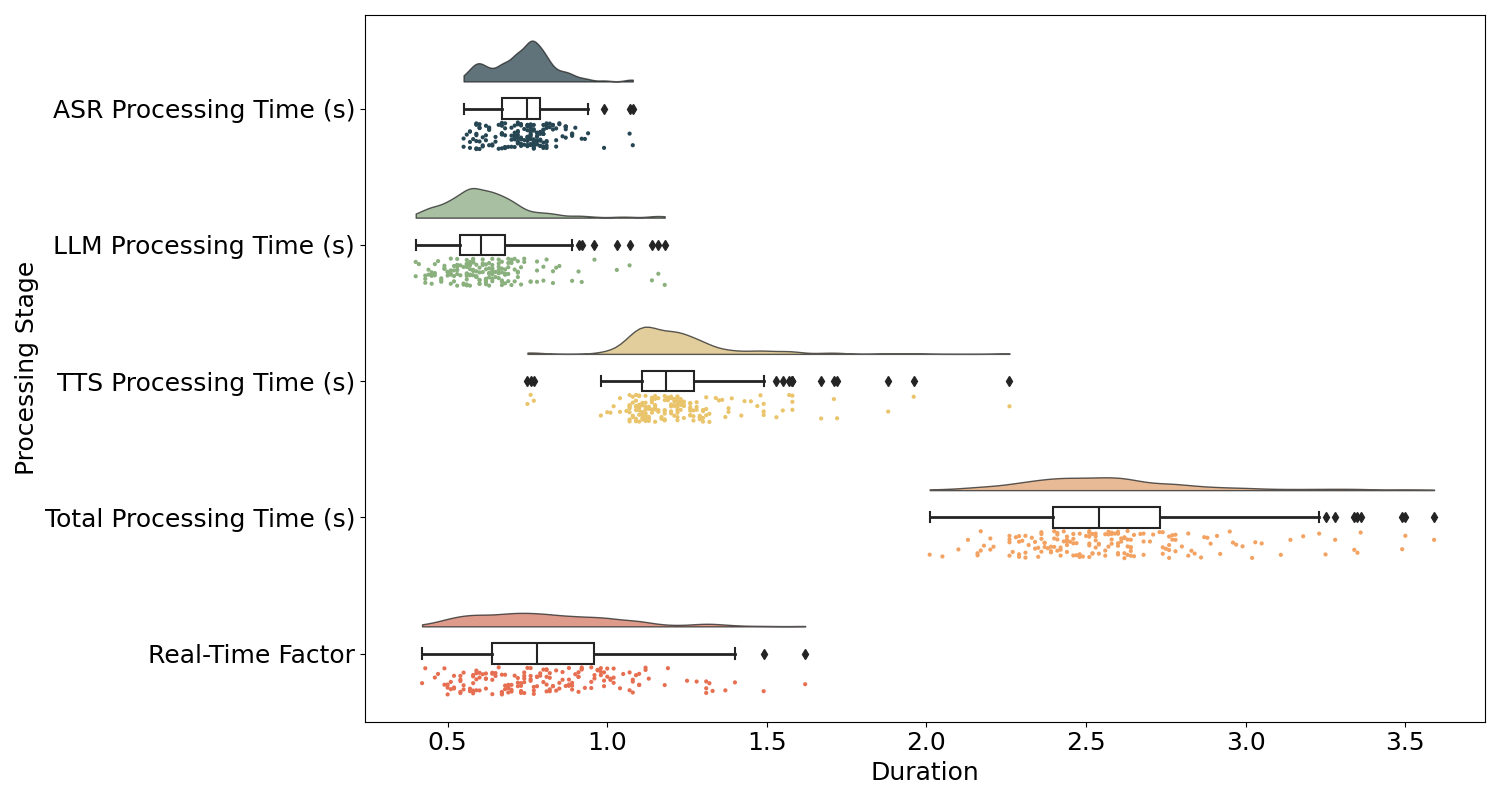}
\caption{Distribution of Processing Times Across Different Stages} \label{fig_latecy}
\end{figure}

Performance analysis demonstrated that Re-Sonance achieves a Real-Time Factor of 0.8189 (Mdn = 0.7800, SD = 0.2396), confirming its efficiency and reliability in speech transcription and optimization tasks. The distribution of specific latency parameters is shown in Figure.~\ref{fig_latecy}.

\section{Discussion and Conclusion}

\subsection{Limitations}
This study has several limitations:
First, while CDSD provides diverse dysarthric speech samples, the system may not generalize to all patient profiles, particularly severe cases or those with atypical speech patterns. Our results showed reduced performance for users with more severe dysarthria.
Second, evaluations were conducted solely in Mandarin Chinese. Although Re-Sonance integrates multilingual components (Whisper ASR, Qwen LLM, Cosyvoice), its cross-linguistic effectiveness remains unverified due to potential language-specific dysarthric variations.
Third, despite real-time optimizations, latency from model inference and network transmission limits precise speech synchronization.
\subsection{Potential of LLMs in Correcting Speech Interaction Errors}
LLMs show strong potential in enhancing speech interaction for individuals with dysarthria. Even without ASR fine-tuning, integration of LLMs significantly improved accuracy and comprehensibility through contextual inference and automatic correction.
LLMs not only correct ASR errors but also infer missing semantic content, especially in complex or unclear speech. Effectiveness depends on input quality, model configuration, and prompt design—structured prompts notably improved alignment with user intent.
Objective evaluations confirm that well-optimized prompts enhance correction accuracy and reduce semantic drift, underscoring LLMs’ value in future AAC systems.
\subsection{Future Work and Research Agenda}
This study lays the foundation for a more accessible AAC system and highlights key research directions:
First, improving intelligibility for users with severe dysarthria remains essential. Future work will explore ASR fine-tuning and personalization to address this gap.
Second, longitudinal user studies are needed to assess real-world usage patterns and social impacts, which may not be captured in lab settings.
Third, cross-linguistic validation is necessary. While Re-Sonance supports multiple languages, systematic evaluation across linguistic and cultural contexts is required, including adaptations for language-specific dysarthria traits.
These directions aim to advance universally accessible AAC technologies and promote communication equity for individuals with dysarthria worldwide.


%
%
%
%
\bibliographystyle{splncs04}  
\bibliography{mybibliography} 

\end{document}